\documentclass{webofc}
\usepackage[varg]{txfonts}   
\usepackage[frozencache=true,cachedir=.]{minted}
\usepackage{natbib}
\newcommand{\mpython}[1]{\mintinline{python}|#1|\xspace}
\newcommand{\arrayfunction}[0]{\mpython{__array_function__}}
\newcommand{\heptables}[0]{\mpython{hep_tables}}
\newcommand{\numpy}[0]{\mpython{numpy}}
\newcommand{\awkward}[0]{\mpython{awkward_array}}
\newcommand{\lambdaf}[0]{\mpython{lambda}}
\newcommand{\jaggedarray}[0]{\mpython{JaggedArray}}
\newcommand{\JAX}[0]{\mpython{JAX}}
\newcommand{\dataframe}[0]{\mpython{DataFrame}}
\newcommand{\dataframeexpressions}[0]{\mpython{dataframe_expressions}}
\newcommand{\xAOD}[0]{\mpython{xAOD}}
\newcommand{\servicex}[0]{\texttt{ServiceX}\xspace}
\setminted[python]{breaklines,breakafter=.,fontsize=\footnotesize,frame=lines,linenos,xleftmargin=0.5in}
\begin{document}
\title{\texttt{hep\_tables}: Heterogeneous Array Programming for HEP\footnote{This work was supported by the National Science Foundation under Cooperative Agreement OAC-1836650.}}
%
%

\author{
    \firstname{Gordon} \lastname{Watts}\inst{1}\fnsep\thanks{\email{gwatts@uw.edu}}
}

\institute{
  University of Washington, Seattle
}

\abstract{%
Array operations are one of the most concise ways of expressing common filtering and simple aggregation operations that is the hallmark of the first step of a particle physics analysis: selection, filtering, basic vector operations, and filling histograms. The High Luminosity run of the Large Hadron Collider (HL-LHC), scheduled to start in 2026, will require physicists to regularly skim datasets that are over a PB in size, and repeatedly run over datasets that are 100's of TB's – too big to fit in memory. Declarative programming techniques are a way of separating the intent of the physicist from the mechanics of finding the data, processing the data, and using distributed computing to process it efficiently that is required to extract the plot or data desired in a timely fashion. This paper describes a prototype library that provides a framework for different sub-systems to cooperate in producing this data, using an array-programming declarative interface. This prototype has a \servicex data-delivery sub-system and an \awkward array sub-system cooperating to generate requested data. The \servicex system runs against ATLAS \xAOD data.
}
\maketitle
\section{Introduction}
\label{sec:intro}
A particle physicist uses a number of heterogeneous systems to make a plot~\cite{atlas-computing-model}. A typical workflow might start with reconstructed data - data that contains physics objects - located in an experiment's production system. The output of the experiment's production system is often located on the GRID, a loosely federated collection of CPU and disk farms. The physicist must submit a job to run on the GRID to access that data. This job extracts the data, applies corrections, and writes out a the data in a simplified format, mostly likely ROOT \texttt{TTree}'s. The physicist then downloads the data locally, and uses local tools to run over that data to extract plots. Time scales can be long: while the jobs on the GRID will normally run quickly, GRID site down times, data delivery glitches, etc., often mean there is a long trail when the user runs on 100's of files~cite{grid-performance}. For many analyses this phase takes more than a week. Once the data is stored locally, it can take approximately a day to produce a plot. However, if the physicist decides that they need a new quantity from the original production data, the cycle must be restarted. For this reason, physicists tend to extract a large amount of data, making the local data sets 10's to 100's of TB.

A second aspect makes this harder than, perhaps, it needs to be: the physicist must know multiple programming languages and tools: C++ and python for the GRID jobs, along with command line tools and scripts to submit and babysit the jobs. For making the plots one has to use either C++ and ROOT or Python and sci-kit-hep set of libraries. Further, one has to have more than a passing familiarity with the complexity of distributed computing systems, along with how well the weakest GRID site is working so that it can be avoided.

The python ecosystem has developed a compelling tool-set that uses array-processing semantics to quickly and intuitively analyze that can be loaded as in-memory arrays. In particle physics, the \awkward array library~\cite{awkward-array} is the most popular of these libraries, and it borrows and extends semantics from the famous rectilinear \numpy~\cite{numpy-paper} array processing library to work with non-uniform multi-dimensional arrays (e.g. \jaggedarray's). This project explores using an array-like interface to take over not only the last step in the plot production process, but also the part that uses the GRID. As an example, the following code will make a plot of electron $p_T$ starting from a set of GRID files that are the output of the ATLAS Monte Carlo production:

\begin{listing}[H]
\begin{minted}{python}
from func_adl import EventDataset
from hep_tables import xaod_table, make_local

dataset = EventDataset('localds://mc15_13TeV:mc15_13TeV.361106.PowhegPythia8EvtGen_AZNLOCTEQ6L1_Zee.merge.DAOD_STDM3.e3601_s2576_s2132_r6630_r6264_p2363_tid05630052_00')
df = xaod_table(dataset)

eles = df.Electrons("Electrons")
good_eles = eles[(eles.pt > 50000.0) & (abs(eles.eta) < 1.5)]

np_pts_good_short = make_local(good_eles.pt/1000.0)

plt.hist(np_pts.flatten(), range=(0, 100), bins=50)
plt.xlabel('Electron $p_T$ [GeV]')
_ = plt.ylabel('PowhegPythia8EvtGen_AZNLOCTEQ6L1_Zee')
\end{minted}
\caption{A complete \heptables code snippet that extracts the electron $p-T$ from an ATLAS \xAOD Monte Carlo sample, brings it to the local machine, and uses \mpython{matplotlib} to plot it as a histogram.} 
\label{lst:ele-pt}
\end{listing}

Line 4 declares the GRID dataset to be used, lines 7-8 define a good electron as being central and $p_T>50000$ MeV (the units in an ATLAS \xAOD file). Line 10 triggers the rendering of the data, which involves calling out to the \servicex~\cite{servicex} backend to fetch the data. It is returned as an awkward array, which is plotted in lines 12-14.

\section{Library Structure}
\label{sec:structure}

This library provides a user-interface that looks much like \numpy or \awkward. Instead of immediately executing the operations, however, it records them. When the physicist request the actual plot or other result, that triggers the library to resolve the query into a set of actions via plug-ins. The end-result is a number, or a column of data (rendering of plots is planned).

The library is built in two layers. The first layer is responsible for recording the actions in the code. The second layer is responsible for executing them. The most popular libraries, like \numpy and \awkward, do both these operations at time: an expression like x[0]+y[0] is executed as it is evaluated. Two advantages are gained by splitting this operation in to recording and execution. First, the operations can be fused - so the addition and the array slicing can take place at the same time. This avoids the creation of temporaries. This is done by analyzing the recording and looking for operations that can be fused. Second, the the recording can easily be efficiently shipped to a remote system where it is executed and only the results transmitted back. Or the execution could be split across multiple systems, without the physicist having to track what is going on.

\subsection{The Interface}
\label{sec:interface}

There are standards for array programming in the python eco-system. Two in particular. For recliner arrays, \numpy is the standard, and for jagged arrays, \awkward is the standard. A huge amount of effort has gone into both interfaces and it would be counter productive to alter them. None-the-less, there are certain types of queries that are difficult to encode, for example, associated object relationships.

\paragraph{\numpy} The original array programming interface. Python has standardized some of the low-level interface work in a PEP~\cite{python-array-pep-3118} (python language extension). Defines basic operators, slicing, and accumulation functions. Designed specifically for rectilinear data.

\numpy operations, of course, are expected to be performed on \numpy arrays. In the case of a project like \heptables, one really wants to record the operations for execution later. \numpy provides an interface, \arrayfunction, which is designed for this~\cite{numpy-array-interface}. As a result, a call to \mpython{numpy.abs(a)} is turned into a call to \mpython{a.}\arrayfunction. The object \mpython{a} can then record the fact that \mpython{numpy.abs} was called on it.

\paragraph{awkward} HEP data is not rectilinear, of course, and \awkward was written to extend \numpy to support more complex arrays, called \jaggedarray's. Besides defining slicing across non-uniform array runs, it also defines operations that work on across array axes, for example, counting the number of jets in an event to yield a number-of-jets-per-event column. \awkward also provides other conveniences like behaviors, which allow an end user to imbue an array with additional behaviors (a 4-vector can be given Lorentz-like semantics).

\paragraph{Extending the Interface} While almost any operation desired is possible with these interfaces, there are places where it is quite difficult to encode them - or that they are not at all straight forward. Consider the following code snippet that matches a Monte Carlo particle to a jet using $\eta-\phi$ matching:
\begin{minted}{python}
result = []
for e in electron:
  min_value = 0.1
  particle = None
  for t in truth:
    if dr(e,t) < min_value:
      min_value = dr(e,t)
      particle = t
  result.append(particle)
\end{minted}
A particle physicist will look at this and know almost immediately what is being done: find the closest MC particle to the electron. Array programming, however, is not nearly as straight forward as it operates on complete arrays at once.

\subsection{Dataframe Expressions}
\label{sec:df-expressions}

There is no python library that will tape-record a set of array operations. There are many libraries that implement \numpy semantics and are drop-in replacements~\cite{dask, ray, vaex}. However, their re-implementation of the \numpy semantics is tightly tied to their implementation.

\dataframeexpressions is built to be a standalone library that simply records the operations made to \dataframe. The library supplies an abstract \dataframe type; and then records all operations done to it. It makes as few assumptions about the underlying data model as possible. Operations occur on the \dataframe type, and include:

\begin{itemize}
    \item The basic unary and binary math operations (e.g. $-a$, $a+b$, $a/b$, etc.)
    \item A number of python built-in functions (e.g. $abs(a)$)
    \item Array leaf references and functions (e.g. $a.jets$ and $a.jets.count()$)
    \item numpy type array references (e.g. $numpy.sin(a)$)
    \item Filtering, or array, slicing (e.g. \mpython{a.jets[a.jets.pt > 30]}, $a.jets[abs(a.jets.eta)<2.4]$
\end{itemize}

The recording is done with a combination of a tree structure and a python abstract syntax tree (AST). The python AST is rich - the complete python language can be expressed in it (by design). Each \dataframeexpressions \dataframe object holds a link to a python AST. If $a$ and $b$ are \dataframe objects, then \mpython{a+b} becomes a new \dataframe object, which contains an AST. The AST is the binary \mpython{+} operator, with reference to the \mpython{a} and \mpython{b} \dataframe's. All expressions are encoded in this way, building a tree of operations.

To use the \dataframeexpressions \dataframe, a library creates the base \dataframe. Usually this means defining the source data (a file, etc.). The user then manipulates the \dataframe's, finishing with some sort of a call to render the results of the expression, like the \mpython{make_local} call in the first listing above. At that point, the library takes the \dataframe that represents the final result and can unwind the user's intent, executing it as required.

One advantage of separating the recording of user intent and the rendering of the expression is \dataframeexpressions \dataframe can provide some short cuts. These short cuts then do not need to be directly handled by the back-end library. Filter functions are one example:
\begin{minted}{python}
def good_jet(j):
    return (j.pt > 35) & (abs(j.eta) < 2.5) & j.isGood
    
good_jets = a[good_jet]
\end{minted}
Functions like this are a help when one needs to define a \mpython{good_jet} in multiple places. Behind the scenes, \dataframeexpressions just calls the lambda with a \dataframe that represents the jet, and records the operations performed on it.

Python \lambdaf functions are also supported. For example, \mpython{a.jets.pt/1000.0} and \mpython{a.jets.map(lambda j: j.pt/1000.0)} resolve to identical array expressions (given the interpretation library applies \mpython{map} as you'd expect here). Python \lambdaf functions can be used in most places in the code, but their power comes from lambda capture - where a variable declared in the outer lambda is captured by an inner lambda. For example, calculating the $\Delta R$ between a jet and electrons: \mpython{a.jets.map(lambda j: a.electrons.map(lambda e: j.DeltaR(e)))}. This creates a 2D \jaggedarray for each event. The first axis is the jet, and the second axis is the $\Delta R$ between that row's jet and each electron in the event. It is now possible to find the electron closest to each jet using sorting, or arg-sorting, or indexing.

It is also possible to add new expressions anywhere in the data model. If an analysis group wants to define \mpython{good_jet}'s for everyone, in a header file it would be possible to write:
\begin{minted}{python}
def define_shortcuts(df):
    df['good_jets'] = df.jets[good_jet]
    
define_shortcuts(a)
good_pts = make_local(a.good_jets.pt)
\end{minted}
This provides some level of composability - as you can easily chain complex expressions, and still conveniently define them inside a function without having to return lots of different values. For example, one could define a new property of a jet, which was the closest Monte Carlo parton, using $\eta-\phi$ matching.

One particular place this proves useful is is defining a macro at one level, but using it at another. For example:
\begin{minted}{python}
a.jets['ptgev'] = a.jets.pt / 1000.0

jetpt_in_gev = a.jets[a.jets.ptgev > 30].ptgev
\end{minted}
Note that the \mpython{ptgev} is defined before the filter. But used after the filter. It is intuitive that this should work, of course, but technically it turned out to be non-trivial.

Leaves can also be referred to as strings - which helps a great deal in simplifying automated histogram and plot generation: \mpython{a.jets['pt']} does exactly what you'd expect it to do. The backend library does not see any of these shortcuts - it just gets the fully resolved AST to process.

\subsection{HEP Tables}

While \dataframeexpressions provides the user interface for the package, \heptables is the implementation. Its job is to take a \dataframe and to render it with the data producing a histogram or other final product.

All \dataframeexpressions start with an initial \dataframe object. \heptables defines a single root \dataframe which currently refers to an ATLAS \xAOD dataset (\mpython{xaod_table}). These datasets are registered on the GRID; despite being stored around the world, the dataset names can be thought of a unified namespace. Lines 4-5 in Listing~\ref{lst:ele-pt} do this. The fact that this is restricted to \mpython{func_adl}'s \mpython{EventDataset} is an artifact of the current implementation (see Section~\ref{sec:status}). The \mpython{xaod_table} is directly derived from a \dataframeexpressions \dataframe.

Until the \mpython{make_local} function call in line 10, everything is handled by the \dataframeexpressions package: no actual execution is scheduled. The \mpython{make_local} triggers the execution. The \heptables package uses utilities in the \dataframeexpressions package to build the full AST that summarizes the calculation. It then uses a plug-in system to create a plan for execution.

\begin{figure}[t]
\centering
\includegraphics[width=0.38\textwidth, clip]{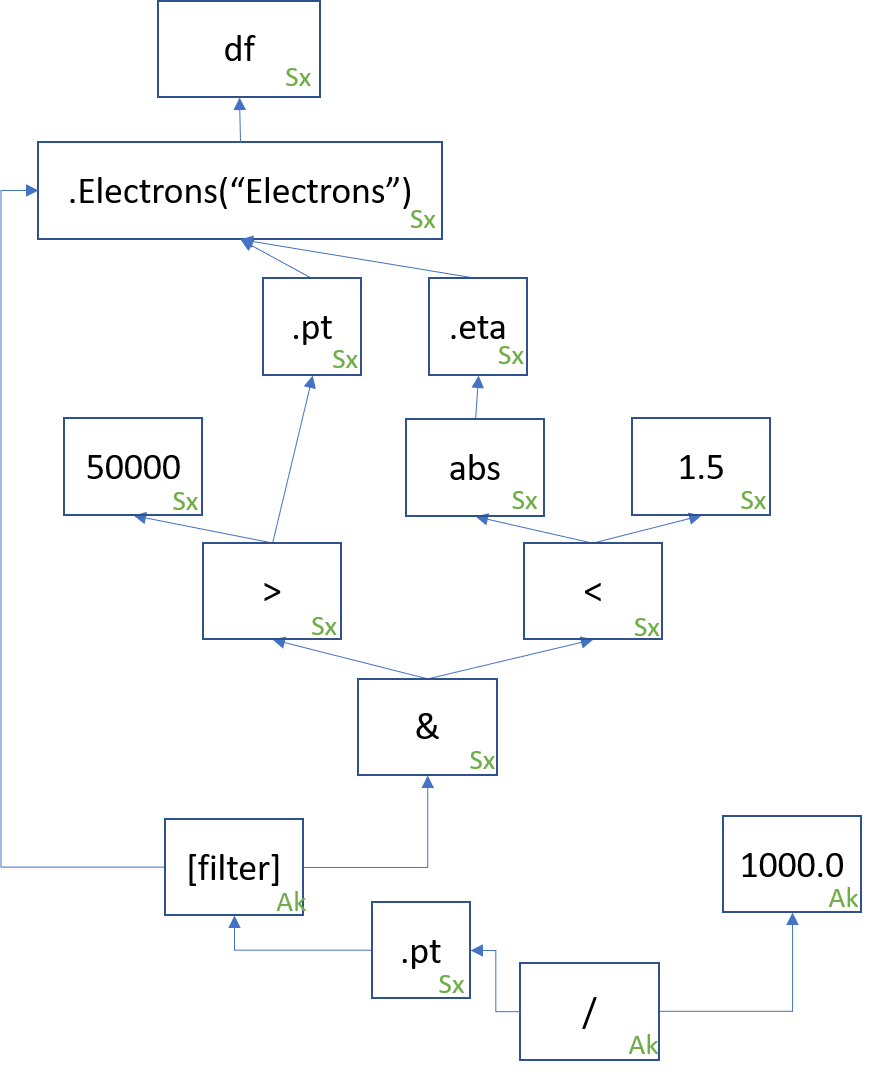}
\caption{The \dataframe tree recorded by the calculation of the good electron $p_T$. The text in each node's lower right indicates which backend plug-in will handle that node's calculation.}
\label{fig:df-frame}
\end{figure}

Currently a \servicex and an \awkward plug-in exist. Each plug-in can look at a single level in the AST and decide if it can execute it (given it can execute all parents). \heptables starts at the deepest parts of the tree and finds a backend that can execute each node of the AST. It does its best to use the executor that could handle all the parents. When the executor switches, \heptables creates a python \mpython{async} task to perform the execution via the selected plug-in. The plug-in's are expected to render \awkward arrays, and the different plug-in's use these to move data between them.

Figure~\ref{fig:df-frame} shows the AST that is produced by the \dataframeexpressions starting from the \mpython{xaod_table} defined by the \heptables package (see Listing~\ref{lst:ele-pt}). Each operation or element is a new \dataframe. \heptables has walked the tree, labeling each node with a plug-in. In this example, all the nodes can be executed by the \servicex plug-in, however due to limitations in the prototype backend, it cannot. When \heptables gets to the bottom of the tree (the filter and division), it will submit to \servicex a request to render the data from the \texttt{\&} and \texttt{.pt} nodes, using the \awkward backend to combine them.

\paragraph{\servicex}

The \servicex plug-in works by translating the AST tree to \mpython{func_adl}, and submitting the query via the \servicex web API. The result is returned as \awkward arrays.

Translation to \mpython{func_adl} for an expression tree like above requires some finesse. The tree contains multiple references to the electron branch, and those that are separated by a distance, the \mpython{.pt} and the \mpython{[filter]} nodes, require an extra carry-along in the \mpython{func_adl} expression. While \mpython{func_adl} has no trouble handling this, the first version of the translator cannot. Though the AST in Figure~\ref{fig:df-frame} could be run entirely in \mpython{func_adl}, it isn't.

\paragraph{\awkward}

The \awkward plug-in is an immediate translator: it executes each AST node, one at a time. It does not try to take advantage of operation fusing. In that sense, this is a direct translation of awkward operations (which also made it very easy to write).

\section{Capabilities}

The capabilities of \heptables are, to first ordered, defined by the backend plug-in's. This section notes a few capabilities that are implemented:

\begin{itemize}
    \item \mpython{map}: The map function is implemented (see Section~\ref{sec:df-expressions}). It takes whatever is to the left as a collection and loops over it. The exact operation depends on what is to the left and the \lambdaf function, of course.
    
    \item Backend-Functions: There are often utility functions declared in the backend. \heptables needs to know about the existence of these, so they must be declared up front. The \mpython{DeltaR} function is an example of this. Any function can be declared, and there is the possibility of shipping arbitrary C++ down to run on the \xAOD/\servicex backend as well, though that is not currently implemented.
    
    \item \mpython{First} and \mpython{Count} and similar functions: \mpython{a.First()} will return the first element of an array, and \mpython{a.Count()} returns number of items in an array. Both of these turn an array into a scalar. Various other \emph{aggregation} functions are also supported.
    
    \item Types are tracked through the expression, though heuristics are used at the start. For example, collections like \mpython{Electrons} are hard-coded into an \xAOD data model. But once the collection is accessed, its type is tracked. The leaves, like \mpython{.pt}, are assumed to be a double, for example. This works out mostly, but not always.
\end{itemize}

\section{Associated Objects Example}

This looks at an example that looks at reconstructed electrons from an \xAOD file, and plots the matching Monte Carlo electron. The match is defined as an electron that is within $\Delta R < 0.1$. It demonstrates a number of the concepts seen above, and illustrates some of the pain points.

\begin{minted}{python}
from hep_tables import xaod_table, make_local, curry
from dataframe_expressions import user_func, define_alias
from func_adl import EventDataset
import matplotlib.pyplot as plt
import numpy as np

@user_func
def DeltaR(p1_eta: float, p1_phi: float, p2_eta: float, p2_phi: float) -> float:
    # Calculate DeltaR between two eta/phi combinations. Implemented in the backend.
    assert False, 'This should never be called'

dataset = EventDataset('localds://mc15_13TeV:mc15_13TeV.361106.PowhegPythia8EvtGen_AZNLOCTEQ6L1_Zee.merge.DAOD_STDM3.e3601_s2576_s2132_r6630_r6264_p2363_tid05630052_00')
df = xaod_table(dataset)

mc_part = df.TruthParticles('TruthParticles')
mc_ele = mc_part[(mc_part.pdgId == 11) | (mc_part.pdgId == -11)]

eles = df.Electrons('Electrons')

def good_e(e):
    'Good electron particle'
    return (e.ptgev > 20) & (abs(e.eta) < 1.4)

good_eles = eles[good_e]
good_mc_ele = mc_ele[good_e]

def associate_particles(source, pick_from):
    def dr(p1, p2):
        'short hand for calculating DR between two particles.'
        return DeltaR(p1.eta(), p1.phi(), p2.eta(), p2.phi())

    def very_near(picks, p):
        'Return all particles in picks that are DR less than 0.1 from p'
        return picks[lambda ps: dr(ps, p) < 0.1]

    source[f'all'] = lambda source_p: very_near(pick_from, source_p)
    
    source[f'has_match'] = lambda e: e.all.Count() > 0
    with_assoc = source[source.has_match]
    with_assoc['mc'] = lambda e: e.all.First()
    
    return with_assoc

matched = associate_particles(good_eles, good_mc_ele)

pt_matched_mc = make_local(matched.mc.ptgev)
pt_matched_reco = make_local(matched.ptgev)

plt.hist((pt_matched_mc-pt_matched_reco).flatten(), bins=100, range=(-20, 20), histtype='step')
plt.ylabel('PowhegPythia8EvtGen_AZNLOCTEQ6L1_Zee')
plt.xlabel("Resolution (truth-reco) [GeV]")
\end{minted}

\begin{figure}
    \centering
    \includegraphics[height=2in]{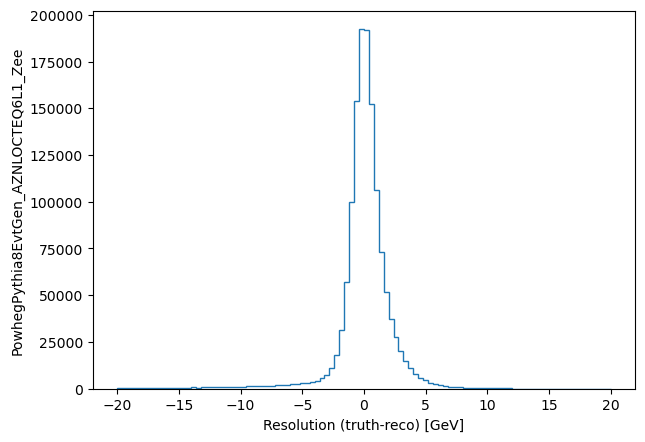}
    \caption{The difference in energy between the reconstructed electron and the MC electron in Monte Carlo}
    \label{fig:ele-diff}
\end{figure}

The result of the plot in line 49 is shown in Figure~\ref{fig:ele-diff}. Some of the code is familiar from Listing~\ref{lst:ele-pt}. Lines 7-10 declare a backend function. In this particular case, that will be executed on \servicex. It is used in the function \mpython{dr} defined at line 28. Lines 24 and 25 define good MC and Data electrons that we'll use in comparison. The function \mpython{associate_particles} defined at line 27 associates the \mpython{source} (electrons) to the \mpython{pick_from} (MC electrons). This is done using the \mpython{very_near} function defined at line 32 that returns  a list of MC electrons that are with in $\Delta R < 0.1$. Note the lambda capture grabs the electron in the form of the variable \mpython{p} that is passed in. Not all electrons have a match, hence the \mpython{has_match} being defined at lien 38, and using it to protect the matched particle definition in line 40. Note that liens 38-40 extend the data model so data electrons now have the \mpython{bool has_match} and an Monte Carlo particle \mpython{mc}. These can be referred to, as they are in line 46, for easy difference plotting.

\section{Status}
\label{sec:status}

A prototype exists in \texttt{github} and is working. This prototype was used to develop the concepts in the code. It dispatches code to run against \servicex and \texttt{awkward} as mentioned earlier. There are, as described, clear short-comings with the code. 
At the time of writing, a new version is being created to fix a number of the architectural mistakes that were made during prototyping.

\begin{itemize}
    \item The \servicex backend is being upgraded to handle the full range of supported \mpython{func_adl} queries to reduce the amount of data that needs to be shipped from \servicex to an analysis cluster.
    
    \item A backend that supports the \texttt{coffea} library is being created. This includes the ability to run processors in multiple places and support for the \texttt{coffea} data model (called \mpython{nanoaod}). One of the biggest benefits here will be the ability to run on clusters. This will replace the \awkward plug-in, as awkward is used by \texttt{coffea}.
    
    \item A full type system is being implemented. If known, this means the system will not have to infer types, and reduce mistakes. It will also mean errors are flagged earlier (e.g. accessing data that doesn't exist).
    
    \item As the users intent is known at the start, it is possible to cache the users query. Thus the second time the same query is run it should take very little time - just a lookup. This is currently implemented by the \servicex backend. We are investigating adding this more generically to the system.
    
    \item Histogram definition and filling is being implemented as something understood by \dataframeexpressions. This will allow histograms (or other similar objects) to be filled in parallel and then combined.
\end{itemize}

\heptables approach, like many in python, does have some issues. Some are a function of this implementation and some are a function of python's programming model.
\begin{itemize}
    \item Assumptions are made about which level an array operation occurs. For example, does \mpython{a.jets.Count()} count the number of jets in each event (returning an array), or count the number of events (returning a scalar)?
    
    \item While \lambdaf capture enables many important patterns, and makes it more clear what is going on, its readability is not the most straight forward. This is a generic problem with python: it is not yet possible to analyze the source code of a python pragmatically in a robust way. The double-loop example in Section~\ref{sec:interface}.
    
    \item It isn't obvious how functions over sequences should work when more than one sequence is involved. For example, \mpython{abs(a.jets.pt)} should return an array of the absolute value of jet pt's. But what about the tuple \mpython{(abs(a.jets.pt), abs(a.jets.pt))}? Should that return a single array, with two entries in each? Or should it return a 2D array?
    
    \item Many common libraries, like \mpython{seaborn}, are used for plotting. This is temptingly close to being able to use these libraries. However, these libraries usually require the data in-memory, and that can be quite expensive for a large analysis dataset.
    
    \item Loop algorithms are not well suited to this style of declarative programming. For example, track finding - where you don't have a definite number of steps.
    
    \item Control logic is not captured - if a decision has to be made on the result of a query, the query must be rendered first. As one thinks forward to adopting differential programming, as the \JAX package~\cite{jaxmd2020}, this problem will have to be better understood.
\end{itemize}

\section{Conclusions}

This paper has described the \dataframeexpressions and \heptables package. The former records the users intent by tracking common and extended array operations. The latter renders the actions across multiple backends. This project was started as a prototype to understand if something like this was possible, without having to re-write the complete eco-system. As both \numpy and \awkward use a dispatch mechanism, this turns out to be possible.

The packages have been used for some simple analysis examples, and are currently undergoing a re-write to make them robust enough to be used for a more sophisticated analysis.

\bibliography{computing-papers}

\end{document}